

\input harvmac


\overfullrule=0pt


\def\bar#1{\overline{#1}}
\def\ccdot{\hbox{\kern-.1em$\cdot$\kern-.1em}}
\def\CD{{\cal D}}
\def\GeV{\>\, \rm GeV}
\def\GF{G_{\scriptscriptstyle F}}
\def\gfive{\gamma^5}
\def\L{{\scriptscriptstyle L}}
\def\LB{{\Lambda_b}}
\def\LBzero{{\Lambda_b^0}}
\def\LC{{\Lambda_c}}
\def\LCplus{{\Lambda_c^+}}
\def\Mb{M_{\LB}}
\def\Mc{M_{\Sigma_c}}
\def\MeV{\> {\rm MeV}}
\def\mpi{m_{\pi}}

\def\slash#1{#1\hskip-0.5em /}
\def\proj{{1 + \slash{v} \over 2}}
\def\pthree{ { \vert \vec{p}_\pi \vert^3 \over f^2}}
\def\space{\>\>}
\def\R{{\scriptscriptstyle R}}
\def\Q{{\scriptscriptstyle Q}}
\def\vv{v \ccdot v'}


\def\a{\alpha}
\def\b{\beta}
\def\e{\epsilon}
\def\g{\gamma}
\def\l{\lambda}
\def\o{\sigma}
\def\u{\mu}
\def\v{\nu}


\def\half{{1 \over 2}}
\def\sixth{{ 1\over 6}}
\def\third{{1 \over 3}}
\def\threehalves{{3 \over 2}}
\def\twothirds{{2 \over 3}}


\newdimen\pmboffset
\pmboffset 0.022em
\def\oldpmb#1{\setbox0=\hbox{#1}%
 \copy0\kern-\wd0
 \kern\pmboffset\raise 1.732\pmboffset\copy0\kern-\wd0
 \kern\pmboffset\box0}
\def\pmb#1{\mathchoice{\oldpmb{$\displaystyle#1$}}{\oldpmb{$\textstyle#1$}}
      {\oldpmb{$\scriptstyle#1$}}{\oldpmb{$\scriptscriptstyle#1$}}}

\def\pib{{\pmb{\pi}}}

\def\LongTitle#1#2#3#4{\nopagenumbers\abstractfont\hsize=\hstitle\rightline{#1}%
\vskip 1in\centerline{\titlefont #2}\centerline{\titlefont #3}
\centerline{\titlefont #4}
\abstractfont\vskip .5in\pageno=0}


\nref\WiseI{M. Wise, Caltech Preprint CALT-68-1721 (1991), Lectures
  presented at the Lake Louise Winter Institute.}
\nref\WiseII{M. Wise, Caltech Preprint CALT-68-1765 (1992).}
\nref\Burdman{G. Burdman and J. F. Donoghue, UMHEP-365 (1992).}
\nref\Grinstein{B. Grinstein, E. Jenkins, A. Manohar, M. Savage and M.
  Wise, UCSD/PTH 92-05 (1992).}
\nref\Yan{T.M. Yan, H.Y. Cheng, C.Y. Cheung, G.L. Lin, Y.C. Lin and H.L.
  Yu, CLNS-92/1138 (1992).}
\nref\CCWZ{S. Coleman, J. Wess and B. Zumino, Phys. Rev. {\bf 177} (1969)
  2239\semi
  C. Callan, S. Coleman, J. Wess and B. Zumino, Phys. Rev. {\bf 177} (1969)
  2247.}
\nref\Manohar{A. Manohar and H. Georgi, Nucl. Phys. {\bf B234} (1984)
  198\semi
  H. Georgi, {\it Weak Interactions and Modern Particle Theory},
  (Benjamin/Cummings, Menlo Park, 1984).}
\nref\Gasser{J. Gasser and H. Leutwyler, Ann. Phys. {\bf 158} (1984) 142\semi
  J. Gasser and H. Leutwyler, Nucl. Phys. {\bf B250} (1985) 465.}
\nref\Eichten{E. Eichten and B. Hill, Phys. Lett. {\bf B234} (1990) 511\semi
  H. Georgi, Phys. Lett. {\bf B240} (1990) 447.}
\nref\GeorgiI{H. Georgi, Heavy Quark Effective Field Theory, HUTP-91-A039
  (1991), Lectures delivered at TASI.}
\nref\Jenkins{
  E. Jenkins and A. Manohar, Phys. Lett. {\bf B255} (1991) 558\semi
  E. Jenkins and A. Manohar, Phys. Lett. {\bf B259} (1991) 353\semi
  E. Jenkins, UCSD/PTH 91-12 (1991)\semi
  E. Jenkins and A. Manohar, UCSD/PTH 91-30 (1991).}
\nref\GeorgiII{H. Georgi, Nucl. Phys. {\bf B348} (1991) 293.}
\nref\Isgur{N. Isgur and M. Wise, Nucl. Phys. {\bf B348} (1991) 276.}
\nref\Randall{H. Georgi and L. Randall, Nucl. Phys. {\bf B276} (1986) 241\semi
  H. Georgi, Nucl. Phys. {\bf B331} (1990) 311.}
\nref\spinorsums{
  H. Umezawa, {\it Quantum Field Theory}, (North-Holland, Amsterdam, 1956)\semi
  T. Mannel, W. Roberts and Z. Ryzak, Nucl. Phys. {\bf B355} (1991) 38\semi
  A. Falk, SLAC-PUB-5689 (1991).}
\nref\Politzer{H. D. Politzer, Phys. Lett. {\bf B250} (1990) 128.}
\nref\PDB{Particle Data Review, Phys. Lett. {\bf B239} (1990) III.45.}
\nref\PDBII{Particle Data Review, to appear in Phys. Rev. {\bf D} suppl.
  (1992).}


\nfig\polegraphs{Leading order pole diagrams that contribute to the
semileptonic process $\LBzero \to \Sigma_c^{++} + e^- + \bar{\nu}_e + \pi^-$.
Strong and weak interaction vertices are denoted by solid circles and squares
respectively.}
\nfig\ratioplot{The dimensionless decay rate ratio $g_3^{-2} R$ defined in
eqn.~\ratio\ plotted as a function of the invariant lepton pair mass
$m_{23}$.}


\LongTitle{HUTP-92/A014}
  {Chiral Perturbation Theory}{for Hadrons Containing a Heavy Quark:}
  {The Sequel}
\centerline{Peter Cho}
\bigskip\centerline{Lyman Laboratory of Physics}
\centerline{Harvard University}
\centerline{Cambridge, MA 02138}

\vskip .3in


	Charm and bottom mesons and baryons are incorporated into a low energy
chiral Lagrangian. Interactions of the heavy hadrons with light octet
Goldstone bosons are studied in a framework which represents a synthesis
of chiral perturbation theory and the heavy quark effective theory.
The differential decay rate for the semileptonic process
$\LBzero \to \Sigma_c^{++} + e^- + \bar{\nu}_e + \pi^-$
is calculated at the zero recoil point using this hybrid formalism.

\Date{3/92}


	Chiral perturbation theory and the heavy quark effective theory
represent two descriptions of hadronic physics that become exact in opposing
limits of QCD \WiseI.  The first is based upon a global
$SU(3)_\L \times SU(3)_\R$
symmetry which is spontaneously broken by the strong interactions to the
diagonal subgroup $SU(3)_{\L+\R}$.  The original chiral and residual flavor
symmetries are only approximate, for they are explicitly violated by quark
masses.  However, since the masses of the three lightest quarks are small
compared to the strong interaction scale $\Lambda_{QCD}$, these
symmetries are reasonably accurate in the real world and are fully restored
in the zero quark mass limit.  The second is derived from an approximate
$SU(6)$ spin-flavor symmetry which results from the masses of the three
heavy quarks in the standard model being large relative to $\Lambda_{QCD}$.
This spin-flavor $SU(6)$ becomes exact in the infinite quark mass limit.

	Both of these effective theories are well-established and
have been widely studied in separate contexts.
Recently however, a synthesis of the two has been proposed
\refs{\WiseII,\Burdman}.  Interactions of heavy mesons with light Goldstone
bosons have been discussed in a chiral Lagrangian framework.  Applications
of this new hybrid formalism to semileptonic $B$ and $D$ decays with
slow pion emission have been considered.
In addition, $SU(3)$ breaking contributions to
heavy meson decay constant ratios as well as $B-{\bar{B}}$ mixing matrix
elements have been analyzed \Grinstein.
In this letter, we incorporate baryons containing a single heavy quark
into this picture and investigate semileptonic transitions
among these hadrons.~\foot{Similar work has recently been reported in
ref.~\Yan.}

	To begin, we briefly review the standard procedure for constructing
low energy chiral Lagrangians \refs{\CCWZ,\Manohar,\Gasser}.  The
Goldstone bosons in the pion octet
$$ \pib = {1 \over \sqrt{2}}
\pmatrix{ \sqrt{\half} \pi^0 + \sqrt{\sixth} \eta & \pi^+ & K^+ \cr
\pi^- & - \sqrt{\half} \pi^0+\sqrt{\sixth}\eta & K^0 \cr
K^- & \bar{K}^0 & - \sqrt{\twothirds}\eta \cr} $$
are first arranged into the exponentiated matrix functions
$\Sigma=e^{2i \pib/f}$ and $\xi=\sqrt{\Sigma}=e^{i \pib/f}.$
The parameter $f \approx 93 \MeV$ that enters into these definitions
corresponds at lowest order to the pion decay constant.
The exponentiated fields transform under the chiral symmetry group as
\eqn\Sigmafields{\eqalign{\Sigma &\to L \Sigma R^\dagger \cr
\xi &\to L \xi U^\dagger = U \xi R^\dagger \cr}}
where $L$ and $R$ represent global $SU(3)_\L$ and $SU(3)_\R$ transformations.
The matrix $U$ is a complicated nonlinear function of $L$,$R$
and $\pib$ which acts like a local transformation under the
diagonal flavor subgroup.  Chiral invariant terms can then be built up from
the fields in \Sigmafields\ and their derivatives.  To leading order in a
derivative expansion, the phenomenological Lagrangian that describes the self
interactions of the Goldstone bosons is simply
\eqn\Lpionzero{\CL^{(0)} = {f^2 \over 4} \Tr(\partial^\u
\Sigma^\dagger \partial_\u \Sigma) + {f^2 \over 2} \Tr(\Sigma^\dagger \u M
+ \u M^\dagger \Sigma).}
Explicit chiral and flavor symmetry breaking effects are represented
in this Lagrangian by the constituent mass parameter $\u$ and the current
quark mass matrix
$$ M = \pmatrix{m_u && \cr & m_d & \cr && m_s \cr}. $$

	Meson and baryon matter fields can generally be included into
the effective Lagrangian.  Their interactions with the pion octet are
governed solely by their light flavor symmetry properties.  The Goldstone
bosons couple derivatively to matter fields through the vector and axial
vector combinations
\eqna\pioncurrents
$$ \eqalignno{
{\bf V}^\u &= \half (\xi^\dagger \partial^\u \xi
  + \xi \partial^\u \xi^\dagger)
  = {1 \over 2f^2} [ \pib, \partial^\u \pib]-{1 \over 24 f^4} \Bigl[ \pib,
  \bigl[\pib, [\pib,\partial^\u \pib] \bigr] \Bigr] + O(\pib^6)
  \qquad & \pioncurrents a \cr
{\bf A}^\u &= {i \over 2}(\xi^\dagger \partial^\u \xi - \xi \partial^\u
  \xi^\dagger) = -{1 \over f} \partial^\u \pib + {1 \over 6 f^3}
  \bigl[\pib,[\pib,\partial^\u \pib] \bigr] + O(\pib^5).
  & \pioncurrents b\cr} $$
The vector acts like an $SU(3)_{\L+\R}$ gauge field
$$ {\bf V}^\u  \to U {\bf V}^\u U^\dagger + U \partial^\u U^\dagger $$
while the axial vector simply transforms as an $SU(3)_{\L+\R}$ octet:
$$  {\bf A}^\u \to U {\bf A}^\u U^\dagger. $$

	We would specifically like to incorporate hadrons that contain a single
heavy quark $Q$.  Following the approach developed for the heavy quark
effective theory \Eichten, we work with velocity dependent fields whose
interactions are constrained by an $SU(2)_v$ spin symmetry group.
We start with the operators $P$ and $P^*_\u$ that annihilate $J^P= 0^-$ and
$1^-$ mesons with quark content $Q \bar{q}$.  If the heavy quark constituent
is charm, the individual components of these fields are
$$ \vbox{\settabs\+
  \hfil $P^*_1$ & $=$ & ${D^*}^0$ \hfil & \qquad &
  \hfil $P^*_2$ & $=$ & ${D^*}^+$ \hfil & \qquad &
  \hfil $P^*_3$ & $=$ & ${D^*_s}^+$ \hfil & \qquad \cr
\+ $P_1$ & $=$ & $D^0$ && $P_2$ & $=$ & $D^+$ && $P_3$ & $=$ & $D^+_s$ \cr
\+ $P^*_1$ & $=$ & ${D^*}^0$ && $P^*_2$ & $=$ & ${D^*}^+$ && $P^*_3$ & $=$ &
   ${D^*_s}^+$ . \cr} $$
The pseudoscalar and vector meson operators can be combined into the
$4\times 4$ matrices \refs{\WiseII,\GeorgiI}
\eqn\Hfield{\eqalign{
H_i(v) &= \proj \bigl[ - P_i \gfive + P^*_{i \u} \gamma^\u \bigr] \cr
\bar{H}^i(v) &= \bigl[ {P^\dagger}^i \gfive + {P^*_\u}^i \g^\u \bigr] \proj.
  \cr }}
$H$ transforms as an antitriplet matter field under $SU(3)_{\L+\R}$
$$ H_i \to H_j (U^\dagger)^j_i $$
and as a doublet under $SU(2)_v$:
$$ H \to e^{i \vec{\e}\cdot \vec{S}_v} H. $$
The spin symmetry rotates the $P$ and $P^*_\u$ operators in
\Hfield\ into one another.

	We also include baryons with quark content $Qqq$ into the chiral
Lagrangian.~\foot{Chiral perturbation theory for baryons containing no
heavy quarks has been thoroughly studied in ref.~\Jenkins.  Many of the
static fermion techniques described in that body of work are similar to
those used here.}
The light degrees of freedom inside these hadrons carry either
one or zero units of angular momentum.  In the former case, the resulting
$J^P=\half^+$ and $J^P=\threehalves^+$ baryons are degenerate in the
infinite quark mass limit and can be assembled into the matrices \GeorgiII
\eqn\Sfield{\eqalign{
S^{ij}_\u(v) &= \sqrt{\third} (\g_\u+v_\u) \gfive \proj B^{ij}
  + \proj {B^*_\u}^{ij} \cr
\bar{S}_{ij}^\mu(v) &= - \sqrt{\third} \bar{B}_{ij} \proj \gfive
 (\g^\u+v^\u)+ {\bar{B}^{\, *}_{ij}}^\u \proj . \cr}}
The $S$ field obeys the constraints $v^\u S^{ij}_\u = 0$ and
$\slash{v} S^{ij}_\u = S^{ij}_\u$.  It transforms as an
$SU(3)_{\L+\R}$ sextet
$$ S^{ij}_\u \to U^i_{i'} U^j_{j'} S^{i'j'}_\u$$
and is an axial vector under parity.  When the heavy quark is
taken to be charm, the components of the Dirac spinor operators $B^{ij}$ in
eqn.~\Sfield\ are the $J^P=\half^+$ baryons
\eqn\sextetcomp{\eqalign{
B^{11} &= \Sigma^{++}_c \qquad B^{12}=\sqrt{\half} \Sigma^+_c \qquad
  B^{22} = \Sigma^0_c \cr
& \quad B^{13} = \sqrt{\half} {\Xi^+_c}' \qquad B^{23} = \sqrt{\half}
  {\Xi^0_c}' \cr
& \space\quad\qquad\qquad B^{33} = \Omega^0_c. \cr}}
Their spin-$\threehalves$ counterparts appear in the Rarita-Schwinger field
${B^*_\u}^{ij}$ which satisfies $\g^\u~B^*_\u~=~0$.  These different spin
components are transformed into one another under the action of $SU(2)_v$:
$$ S_\u \to e^{i \vec{\e} \cdot \vec{S}_v} S_\u. $$

	The remaining heavy baryons whose light spectator degrees of
freedom have zero angular momentum are assigned to the matrix
\eqn\Tfield{T_i(v) = \proj B_i}
which is an $SU(3)_{\L+\R}$ antitriplet:
$$ T_i \to T_j (U^\dagger)^j_i.$$
Its conjugate field is simply
$$ \bar{T}^{\, i} (v) = \bar{B}^{\, i} \proj. $$
The components of the singly charmed $B_i$ operators are the $J^P=\half^+$
baryons~\foot{In the absence of a universally accepted nomenclature
convention for
distinguishing between the isospin-$\half$ $\Xi_\Q$ states in the sextet and
antitriplet multiplets, we have followed ref.~\Isgur\ and denoted
the heavier sextet states with a prime.}
$$ B_1 = \Xi_c^0  \qquad B_2 = - \Xi_c^+ \qquad B_3 = \LCplus.$$
The $SU(2)_v$ symmetry rotates the spins of these baryons
which come entirely from their heavy quark constituents:
$$ T \to e^{i \vec{\e} \cdot \vec{S}_v} T. $$

	Before displaying the heavy hadron contributions to the low energy
Lagrangian, we should mention
the power counting rules that can be used to estimate the sizes of their
coefficients \refs{\Manohar,\Randall}.
Each term begins proportional to $f^2 \Lambda^2$ and has the factors
\settabs\+\indent& $\displaystyle{\sqrt{M} / f \sqrt{\Lambda}}$ \quad & \cr
\+ & \hfill $\displaystyle{1 / f}$ \hfill&
   for each strongly coupled light boson, \cr
\+ & \hfill $\displaystyle{\sqrt{M} / f \sqrt{\Lambda}}$ \hfill &
   for each strongly coupled boson containing a heavy quark \cr
\+&& \quad of mass $M$, \cr
\+ & \hfill $\displaystyle{1 / f \sqrt{\Lambda}}$ \hfill &
   for each strongly coupled light fermion, \cr
\+ & \hfill $\displaystyle{1 / \Lambda}$ \hfill &
   for each derivative or dimension one symmetry breaking term. \cr\noindent
Here $\Lambda \approx 4 \pi f \approx 1 \GeV$ represents the chiral symmetry
breaking scale.
The mass of a heavy meson can be substituted in place of the mass of its
heavy quark at lowest order.   All dependence upon heavy meson masses can
subsequently be removed from the zeroth order
Lagrangian via the redefinition $H'=\sqrt{M_H} H$.
The meson field $H'$ then has mass dimension $\threehalves$ like the
$S$ and $T$ baryon fields.

	We can now write down all the leading order terms in the chiral
Lagrangian which are hermitian, Lorentz invariant, parity even,
symmetric under heavy quark spin $SU(2)_v$ and light flavor $SU(3)_{\L+\R}$,
and baryon number conserving:
\eqn\Lvzero{\eqalign{
\CL_v^{(0)} = \sum_{\buildrel Heavy \over {\scriptscriptstyle Flavors}} \Bigl\{
& -i\Tr \bigl( {\bar{H}'}^i v \cdot \CD H_i' \bigr)
  -i \bar{S}^\u_{ij} v\cdot\CD S^{ij}_\u
  +i \bar{T}^i v \cdot \CD \, T_i \cr
& + g_1 \Tr \bigl( H_i' (\slash{{\bf A}})^i_j \gfive {\bar{H}'}^j \bigr)
  + i g_2 \varepsilon_{\u\v\o\l} \bar{S}^\u_{ik} v^\v (A^\o)^i_j (S^\l)^{jk}
  \cr
& + g_3 \Bigl[ \e_{ijk} \bar{T}^i (A^\u)^j_l S^{kl}_\u
    + \e^{ijk} \bar{S}^\u_{kl} (A_\u)^l_j T_i \Bigr] \Bigr\} . \cr}}
A few points should be noted.  Firstly, the matter
field covariant derivatives are constructed from the Goldstone boson vector
current in \pioncurrents{a}:
$$ \eqalign{
\CD^\u H'_i &= \partial^\u H'_i - H'_j (V^\u)^j_i \cr
\CD^\u S^{ij}_\v &= \partial^\u S^{ij}_\v + (V^\u)^i_k S^{kj}_\v
  +(V^\u)^j_k S^{ik}_\v \cr
\CD^\u T_i &= \partial^\u T_i - T_j (V^\u)^j_i. \cr} $$
Partial derivatives acting on the velocity dependent fields which are only
slightly off-shell yield small residual momenta.
Secondly, the signs in front of the kinetic terms have been chosen so that
the meson and baryon components of $H'$,$S$ and $T$ are
conventionally normalized.  Notice that the
spin-$\threehalves$ Rarita-Schwinger fields inside $S^\u_{ij}$ enter
into the kinetic part of \Lvzero\ with opposite sign to their spin-$\half$
counterparts.  Thirdly, we have neglected the mass difference between the
sextet and antitriplet multiplets in this zeroth order Lagrangian.  The
mass difference is phenomenologically comparable to the mass splittings
within the multiplets.  We consequently regard it as a small correction that
should be included with $SU(3)$ breaking effects at next-to-leading order.

	Finally, observe that there is no axial vector term for the
antitriplet baryons like those for the mesons and sextet baryons in
Lagrangian \Lvzero.  Candidate terms such as
$\bar{T} \slash{A} \gfive T$ or $\bar{T} v \ccdot A \gfive T$ either break
the spin symmetry or vanish.
One can understand why such an axial vector interaction cannot exist
by considering a representative process which it would mediate:
$$ \vbox{\settabs\+ \hfil $S_{\rm heavy}$ &:& \qquad \hfil
  & \hfil $\Lambda_\Q$ & $=$ & $Q(qq)$ \quad \hfil & $\to$ \quad
  & \hfil $\Lambda_\Q$ & $=$ & $Q(qq)$ \quad \hfil & $+$ \quad
  & \hfil $\eta$ \hfil \cr
\+ &&&  $\Lambda_\Q$ & $=$ & $Q(qq)$ & $\to$ & $\Lambda_\Q$ & $=$ & $Q(qq)$
     & $+$ & $\eta$ \cr
\+ $S_{\rm heavy}$ &:& && $\half$ &&&& $\half$ &&& $0$ \cr
\+ $S_{\rm light}$ &:& && $0$     &&&& $0$     &&& $0$ \cr
\+             $P$ &:& && $+$     &&&& $+$     &&& $-$ \cr } $$
As a reminder, we have indicated the spins of the heavy quark and the
residual light degrees of freedom as well as the intrinsic parities of the
hadrons involved in this transition.  In order to conserve
angular momentum, the
outgoing hadrons must emerge in an S-wave.  But then the parity of the final
state does not equal the parity of the initial state. So this hadronic process
cannot take place in the infinite quark mass limit of QCD.

	Feynman rules can be simply derived from the
effective Lagrangian.  The Dirac and Rarita-Schwinger spinor sums
\spinorsums
$$ \eqalign{
\Lambda_+ &= \sum_{s=1}^2 u(v,s) \bar{u}(v,s) = \proj \cr
\Lambda_+^{\u\v} &= \sum_{s=1}^4 \CU^\u(v,s) \bar{\CU}^\v(v,s) =
  \bigl[ -g^{\u\v}+v^\u v^\v+\third (\g^\u+v^\u)(\g^\v-v^\v) \bigr]
  \proj \cr} $$
along with the polarization sum
$$ \Lambda^{\u\v} = \sum_{s=1}^3 \e^\u(v,s) \e^\v(v,s)^* = -g^{\u\v}
  + v^\u v^\v $$
appear in the spin-$\half$ and spin-$\threehalves$ baryon and vector meson
propagators $i\Lambda_+/(v\ccdot k)\,$, $i\Lambda_+^{\u\v}/(v\ccdot k)$
and $i\Lambda^{\u\v}/(2v\ccdot k)$ where $k$ denotes the heavy particles'
residual momenta.  Interaction vertices are
established by expanding the velocity dependent fields and Goldstone boson
currents in \Lvzero.  With the Feynman rules in hand, one may readily compute
rates for strong interaction decays of heavy hadrons
with single pion emission.   Some representative examples are listed below:
$$ \eqalign{
\Gamma({D^+}^* \to D^0 \pi^+) &= {g_1^2 \over 12\pi} \pthree \cr
\Gamma( {\Sigma^{++}_c}^* \to \Sigma^+_c \pi^+) &= {g_2^2 \over 72\pi}
  \pthree \cr
\Gamma( {\Sigma^{++}_c}^* \to \Lambda^+_c \pi^+)  &= \Gamma(\Sigma^{++}_c
\to \Lambda^+_c \pi^+) = {g_3^2 \over 12 \pi} \pthree . \cr} $$
In principle, these rates fix the three independent couplings
$g_1$, $g_2$ and $g_3$.  However, the parameters' values cannot yet be
determined given current experimental data.  They are expected to be of
order one.

	Weak semileptonic $b \to c$ transitions can also be investigated
in this chiral Lagrangian framework.  Such
processes are governed by the underlying four-fermion interaction
$$ \CL_{\rm weak} = {4 \GF \over \sqrt{2}} V_{cb} \sum_{\ell = e,\mu,\tau}
  (\bar{\ell} \g^\u P_- \v_\ell) \, (\bar{c} \g_\u P_- b) $$
where $P_- = \half (1-\gfive)$ denotes a left-handed projection operator.
The hadronic current that enters into this weak vertex matches
at zeroth order onto an effective current in the low energy theory which
is specified in terms of four Isgur-Wise functions
\refs{\WiseII,\GeorgiII}:
$$ \eqalign{\bar{c} \g_\u P_- b \to C_{cb} \Bigl\{
&- \xi(\vv) \Tr \bigl( \bar{H}'_c(v') \g_\u P_- H'_b(v) \bigr) \cr
&- \bigl[ g_{\a\b} \eta_1(\vv) - v_\a v'_\b \eta_2(\vv) \bigr]
   \bar{S}^\a_c(v') \g_\u P_- S^\b_b(v) \cr
&+ \eta(\vv)\bar{T}_c(v') \g_\u P_- T_b(v) \Bigr\}. \cr} $$
Perturbative QCD scaling corrections are absorbed into the prefactor
$C_{cb}$.  When $v=v'$, the functions $\xi$,
$\eta_1$ and $\eta$ equal unity while all dependence upon the remaining
$\eta_2$ function disappears.  We will confine our attention to the
kinematic neighborhood around the zero recoil point in order to take
advantage of this tremendous simplification.

	As an illustration of the utility of chiral perturbation theory for
hadrons containing a heavy quark, we consider $\LBzero$ semileptonic decays.
Such processes are
of phenomenological interest since they are among the more readily
identifiable bottom baryon transitions that will be measured
in the future.  We are interested in studying generalizations of the pure
semileptonic decay
\eqn\puresemi{\LBzero(P; v) \to \LCplus(p_1;v) + e^-(p_2) +
  \bar{\nu}_e(p_3)}
that have low momentum Goldstone bosons in the final state.  The simplest
possibility
$$ \LBzero(P;v) \to \LCplus(p_1;v) + e^-(p_2) + \bar{\nu}_e(p_3)
  + \eta(p_4) $$
does not occur at lowest order due to the absence of an axial vector
coupling to the antitriplet baryons.  This process is mediated by
$O(1/M_\Q)$ operators which break the heavy quark spin symmetry.  But
predictive power is diminished at next-to-leading order since those
operators' coefficients are unknown.

	We consider instead the alternative
\eqn\semipion{\LBzero(P;v) \to \Sigma_c^{++}(p_1;v) +
  e^-(p_2) + \bar{\v}_e(p_3) + \pi^-(p_4).}
The corresponding transition with no final state pion violates both isospin
and strong parity of the light degrees of freedom within the heavy hadrons
\refs{\Isgur,\Politzer}.  Therefore, decay \semipion\ most likely
represents the dominant $\LBzero \to \Sigma_c^{++}$ semileptonic mode.  It
proceeds via the three pole diagrams illustrated in \polegraphs.
Adding these graphs together, squaring the resulting amplitude,
and averaging and summing over fermion spins, we obtain the total squared
amplitude
\eqn\squareamp{\eqalign{\half \sum_{\rm spins} |\CA|^2 &=
  - {16 \over 27} \GF^2 \Bigl({g_3 \over f}\Bigr)^2 |V_{cb}|^2 C_{cb}^2 \cr
&\quad \times \Bigl\{ 5\bigl[ p_2 \ccdot p_4 \, p_3 \ccdot p_4
  -  p_2 \ccdot p_4 \, v\ccdot p_3 \, v\ccdot p_4
  -  p_3 \ccdot p_4 \, v\ccdot p_2 \, v\ccdot p_4
  +  v \ccdot p_2 \, v\ccdot p_3 \, (v\ccdot p_4)^2 \bigr] \cr
&\qquad\qquad + \bigl[ 2 p_2\ccdot p_3 + 9 v\ccdot p_2 \, v\ccdot p_3 \bigr]
  \bigl[ p_4^2 - (v\ccdot p_4)^2 \bigr] \Bigr\}/ (v \ccdot p_4)^2. \cr}}
The differential rate for \semipion\ is then given by \PDB
\eqn\diffdecay{d\Gamma =  {1 \over 8} (2\pi)^{-8} {M_{\Sigma_c} \over \Mb}
\Bigl( \half \sum_{\rm spins} |\CA|^2 \Bigr) |{\vec p}_1| \, d\Omega_1 \,
|\vec{p}\, '_4| \, d\Omega'_4 \, dm_{234} \,
|\vec{p}\, ''_2| \, d\Omega''_2 \, dm_{23}.}
In this expression, $m_{23}= \sqrt{p_{23}^2} = \sqrt{(p_2+p_3)^2}$ is
the invariant mass of the lepton pair, and $(|\vec{p}\,''_2|, d\Omega''_2)$
stands for the electron's three-momentum in the rest frame of its virtual
$W^*$ progenitor.  Similarly, $m_{234}=\sqrt{(p_2+p_3+p_4)^2}$, and
$(|\vec{p}\,'_4|, d\Omega'_4)$ represents the emitted pion's momentum in the
$W^*$ and $\pi^-$ center of mass frame.  Finally, $(|\vec{p}_1|, d\Omega_1)$
denotes the momentum of the recoiling $\Sigma_c^{++}$ in the $\LBzero$
rest frame which vanishes of course at the zero recoil point.

	After boosting to the primed and doubleprimed frames to evaluate the
dot products in \squareamp\ and performing the angular integrations in
\diffdecay, we find for the differential width
$$ \eqalign{ {1 \over |\vec{p}_1|} & {d\Gamma(\LB \to \Sigma_c e \bar{\nu}\pi)
  \over dm_{234} \, dm_{23}} \Bigg|_{m_{234} = \Mb - \Mc}  \cr
\qquad & = {2 \over 81} (2\pi)^{-5} \GF^2 |V_{cb}|^2 C_{cb}^2
  \Bigl( {g_3 \over f} \Bigr)^2 {\Mc \over \Mb} {m_{23} \over \Mb - \Mc} \cr
\qquad & \qquad \times { \bigl[ (\Mb-\Mc+\mpi)^2-m_{23}^2 \bigr]
 \bigl[ (\Mb-\Mc-\mpi)^2-m_{23}^2 \bigr] \over
 \bigl[ (\Mb-\Mc)^2+\mpi^2-m_{23}^2 \bigr]^2 } \cr
\qquad & \qquad \times { \sqrt{\bigl[ (\Mb-\Mc)^2-\mpi^2 \bigr]^2
  -2\bigl[ (\Mb-\Mc)^2+\mpi^2 \bigr] m_{23}^2 + m_{23}^4} \over (\Mb-\Mc)^2}\cr
\qquad &\qquad \times \Bigl\{ \bigl[(\Mb-\Mc)^2-\mpi^2 \bigr]^2
  +\bigl[20(\Mb-\Mc)^2-2\mpi^2 \bigr] m_{23}^2+m_{23}^4 \Bigr\}. \cr} $$
For comparison purposes, we normalize this result to the corresponding zero
recoil rate for the pure semileptonic process in \puresemi:
$$ {1 \over |\vec{p}_1|} {d\Gamma(\LB \to \LC e \bar{\nu}_e) \over dm_{23}}
  \Bigg|_{m_{23} = \Mb - M_{\LC}}  = 2 (2\pi)^{-3} \GF^2 |V_{cb}|^2 C_{cb}^2
  {M_\LC \over \Mb} (\Mb - M_\LC)^3.$$
The dimensionless ratio of these two differential decay rates
\eqn\ratio{R = (\Mb-\Mc-\mpi) { {d\Gamma(\LB \to \Sigma_c e \bar{\nu}_e
  \pi) / dm_{234} dm_{23} \Big|_{m_{234} = \Mb - \Mc}} \over
  {d\Gamma(\LB\to \LC e \bar{\nu}_e)/dm_{23} \Big|_{m_{23}=\Mb - M_{\LC}}}
  }}
is plotted in \ratioplot\ as a function of the invariant lepton pair mass
over its range $0 \le m_{23} \le m_{234}-m_4 = \Mb-\Mc-\mpi$.~\foot{We use
the heavy hadron mass values $M_{\LC} = 2285 \MeV$, $\Mc = 2453 \MeV$ and
$\Mb = 5640 \MeV$ \PDBII.}
Since the derivative expansion breaks down as the outgoing pion's momentum
approaches the chiral symmetry breaking scale, the plot can only be trusted
near the high end of the $m_{23}$ range where the leptons carry away most
of the released energy.  However, one can see from the figure
that the rates for the $\LBzero$ semileptonic decays with and without final
state pion emission are comparable.

	Other interesting interactions between very heavy and very light
hadrons can be studied using the hybrid chiral Lagrangian
formalism.  Some questions which cannot be answered by either chiral
perturbation theory or the heavy quark effective theory alone may be
addressed by their union.  The synthesis of the two effective theories
therefore broadens the scope of QCD phenomena that can be sensibly
investigated.

\bigskip
\centerline{\bf Acknowledgements}
\bigskip

	Helpful discussions with Eric Carlson, Howard Georgi, Liz Simmons,
Mark Wise and Tung-Mow Yan are gratefully acknowledged.  I am especially
indebted to Mark Wise for communicating his results prior to publication and
for bringing ref.~\Yan\ to my attention.  (As indicated by the title, this
letter is intended to be a close follow-on to Wise's original work
\WiseII.)  I would also like to thank Tung-Mow Yan for kindly communicating
ref.~\Yan.  Finally, I am grateful to Charles Wohl for providing access to
heavy baryon Particle Group data.

	This work was supported in part by the National Science Foundation
under contract PHY-87-14654 and by the Texas National Research Commission
under Grant \# RGFY9106.

\listrefs
\listfigs
\bye